\documentclass{PoS}

\usepackage{bm,amsmath}
\newcommand{\smallfrac}[2]{\genfrac{}{}{0pt}{1}{#1}{#2}}

\title{Perturbative calculations for the HISQ action: the gluon action at $O(N_f\alpha_sa^2)$}

\ShortTitle{Perturbative calculations for the HISQ action}

\author{A. Hart$^a$, \speaker{G.M. von Hippel}$^b$, R.R. Horgan$^c$\\
        \llap{$^a$} SUPA, School of Physics and Astronomy,
                    University of Edinburgh,
                    Edinburgh EH9 3JZ, U.K.\\
        \llap{$^b$} Deutsches Elektronen-Synchrotron DESY,
                    15738 Zeuthen, Germany\\
        \llap{$^c$} DAMTP, CMS, University of Cambridge,
                    Cambridge CB3 0WA, U.K.\\ \\
        Email: \email{georg.von.hippel@desy.de} }

\abstract{We present a new (and general) algorithm for deriving
          lattice Feynman rules which is capable of handling actions
          as complex as the Highly Improved Staggered Quark (HISQ) action.
          This enables us to perform a perturbative calculation of the
          influence of dynamical HISQ fermions on the perturbative
          improvement of the gluonic action in the same way as we have
          previously done for asqtad fermions. We find the fermionic
          contributions to the radiative corrections in the
          L\"uscher-Weisz gauge action to be somewhat larger for HISQ
          fermions than for asqtad.}

\FullConference{The XXVI International Symposium on Lattice Field Theory \\
                July 14 - 19, 2008\\
                Williamsburg, Virginia, USA}

\begin{document}

%%%%%%%%%%%%%%%%%%%%%%%%%%%%%%%%%%%%%%%%%%%%%%%%%%%%%%%%%%%%%%%%%%%%%%%%%%%

\section{Introduction}

Continuing rapid advances in parallel computing, along with
theoretical progress in the formulation of lattice field theories
with fermions, have led to lattice QCD simulations with dynamical
light quarks becoming the norm rather than the exception.

The Fermilab Lattice, MILC and HPQCD collaborations have
an ambitious program which to date has made several high-precision
predictions from unquenched lattice QCD simulations
\cite{Davies:2003ik}.
This body of work is based on the Symanzik-improved
staggered-quark formalism, specifically the use of the asqtad
\cite{Orginos:1999cr}
action. More recently, the Highly Improved Staggered Quark (HISQ) action
has been used to further suppress taste-changing interactions and to
allow the use of heavier quarks at the same lattice spacing by removing
tree-level $O((ma)^4)$ artifacts from the quark action
\cite{Follana:2006rc}.
In order to consistently use the HISQ action for the sea quarks as well
\cite{Wong:2007uz},
the calculation of HISQ quark loops on the Symanzik-improvement of the
gluon action is also needed. Having previously carried out that calculation
for the asqtad action
\cite{Hao:2007iz},
we update our calculation here to apply to the case of dynamical HISQ fermions.

%%%%%%%%%%%%%%%%%%%%%%%%%%%%%%%%%%%%%%%%%%%%%%%%%%%%%%%%%%%%%%%%%%%%%%%%%%%

\section{Perturbation Theory for the HISQ action}

The HISQ action is defined by an iterated smearing procedure with
reunitarisation:
\begin{equation}
U^{HISQ} = ( F_{\textrm{asq}'} \circ P_{U(3)} \circ F_{\textrm{Fat7}} )[U]
\end{equation}
where $U=\exp(gA)$ is the unsmeared gauge field, $P_{U(3)}$ denotes the polar
projection onto $U(3)$ (as used in simulations, and \textit{not} $SU(3)$),
and the Fat7 and modified asq smearings are defined in
\cite{Follana:2006rc}.
Straightforward application of the methods from
\cite{Hart:2004bd}
to this action is unfeasible, since the memory requirements for expanding
the action directly into monomials quickly become excessive.
We therefore take advantage of the two-level structure
inherent in the definition of the action and split the derivation and
application of the Feynman rules into two steps.

In the first step, the Feynman rules for the outer layer (the modified
asqtad action) are derived in the same way as previously. We use our HiPPy
python code
\cite{Hart:2004bd}
to expand the asq' action in terms of the Fat7R smeared link
\begin{equation}
U_\mu^{\textrm{Fat7R}}(x) = ( P_{U(3)} \circ F_{\textrm{Fat7}} )[U_\mu] = e^{B_\mu(x+\smallfrac{1}{2}\hat\mu)}
\end{equation}
with a Lie-algebra--valued field $B_\mu$, giving the usual monomials
\begin{equation}\label{eqn:asqfr}
V_r=\frac{g^r}{r!}\sum_i f^{\textrm{asq'}}_{r;i} \bar{\psi}(x_{r;i})B_{\mu_1}(v_{r;i,1})\cdots B_{\mu_r}(v_{r;i,r})\Gamma_{r;i}\psi(y_{r;i})
\end{equation}
To derive the full HISQ Feynman rules, we also need to know the expansion of
$B_\mu$ in terms of the original gauge potential $A_\mu$.
To obtain this, we write the Fat7-smeared link
as\footnote{In the following, we will suppress Lorentz and lattice site indices.}
$F_{\textrm{Fat7}}[U] = M = HV$, where $H^\dag=H$ and $V\in U(3)$. We can now use our
HiPPy expansion routines
\cite{Hart:2004bd}
to obtain an expansion
\begin{equation}
M = c[{\bf 1} + a_\mu*A_\mu + a_{\mu\nu}*(A_\mu*A_\nu) + \ldots]
\end{equation}
where, e.g.
$a_{\mu\nu}*(A_\mu*A_\nu) = \sum_{x,y} a_{\mu\nu}(x,y)A_\mu(x+\smallfrac{1}{2}\hat\mu)A_\nu(y+\smallfrac{1}{2}\hat\nu)$.
Then unitarity of $V$ implies that $R\equiv MM^{\dag}=H^2$ and hence
$V=R^{-1/2}M$ using the expansion
\begin{equation}
R^{-1/2}=(1+(R-1))^{-1/2}=1-\frac{1}{2}(R-1)+\frac{3}{8}(R-1)^2+\ldots
\end{equation}
Rearranging the result as $V=\exp(B)$, i.e.
\begin{equation}
B=\log(V)=(V-1)-\frac{1}{2}(V-1)^2+\ldots
\end{equation}
finally yields the desired expansion of $B$.

Given this, we can now numerically reconstruct the HISQ Feynman rules
for any given set of momenta from eqn. (\ref{eqn:asqfr}) by a convolution
of the asq' Feynman rules of eqn. (\ref{eqn:asqfr}) with the expansion of
$B_\mu$ in terms of $A_\mu$,  summing up all the different ways in which
the gluons $A_\mu$ going into the vertex could have come from the fields
$B_\mu$ appearing in eqn. (\ref{eqn:asqfr}).
Compared to a simple-minded expansion of the HISQ action,
this not only save enough memory to enable the derivation to be performed
in practice, but also leads to a considerable speed-up in many cases.
In particular, we can take advantage of the (anti-)symmetries that the
expansion of $B_\mu$ in terms of $A_\mu$ possesses, allowing us to reduce
the number of contributions we need to take into account when evaluating
Feynman diagrams. In the calculation of the three-gluon vertex for the
``octopus'' diagram (a fermion tadpole with three gluon legs) entering the
three-point function, we are able to omit the contribution from the
expansion of a single $B_\mu$ into three gluons on symmetry grounds.

%%%%%%%%%%%%%%%%%%%%%%%%%%%%%%%%%%%%%%%%%%%%%%%%%%%%%%%%%%%%%%%%%%%%%%%%%%%

\section{On-shell improvement}

The L\"uscher-Weisz action is given by
\cite{Luscher:1984xn}
\begin{equation}\label{eqn:lw_action}
S = \sum_x \Bigg\{
(1-8(c_1+c_2)) \sum_{\mu\not=\nu}\left<1-P_{\mu\nu}\right>
+2 c_1 \sum_{\mu\not=\nu}\left<1-R_{\mu\nu}\right>
+\frac{4}{3} c_2 \sum_{\mu\not=\nu\not=\rho}\left<1-T_{\mu\nu\rho}\right>
\Bigg\}\;,
\end{equation}
where $P$, $R$ and $T$ are the plaquette, rectangle and ``twisted''
parallelogram loops, respectively. The coefficients $c_1$ and $c_2$
need to be determined in order to eliminate the $\mathcal{O}(a^2)$
lattice artifacts.

Given two independent quantities $Q_1$ and $Q_2$ with expansions
\begin{equation}
Q_i = \bar{Q}_i + w_i (\mu a)^2 + d_{ij} c_j (\mu a)^2 +
\mathcal{O}\left((\mu a)^4\right)\;,
\end{equation}
in powers of $(\mu a)$, where $\mu$ is some energy scale, we obtain
the $\mathcal{O}(a^2)$ matching condition
\begin{equation}
\label{eqn:impcond_generic}
d_{ij} c_j = -w_i\;.
\end{equation}
Since this equation is linear, both sides can be decomposed
into a gluonic and a fermionic part; the known gluonic part
\cite{Luscher:1985wf,Snippe:1997ru}
being independent of the fermion action, we will here focus
only on the fermionic part.

At tree-level, there are no fermion loops to consider,
and hence the tree-level coefficients remain unchanged
compared to the quenched case \cite{Luscher:1985wf}. To compute
to one-loop fermionic corrections to the gluon action, we will
follow the same procedure as in the case of the asqtad
action \cite{Hao:2007iz}.

%%%%%%%%%%%%%%%%%%%%%%%%%%%%%%%%%%%%%%%%%%%%%%%%%%%%%%%%%%%%%%%%%%%%%%%%%%%

\section{Twisted boundary conditions}

We work on a four-dimensional Euclidean lattice of length $La$ in the
$x$ and $y$ directions and lengths $L_za,~L_ta$ in the $z$ and $t$ 
directions, respectively, where $a$ is the lattice spacing and $L,L_z,L_t$ are
even integers. In the following, we will employ twisted boundary conditions
in much the same way as in
\cite{Luscher:1985wf,Snippe:1997ru}.
The twisted boundary conditions we use for gluons and quarks are
applied to the $(x,y)$ directions and are given by ($\nu=x,y$)
\begin{equation}
U_\mu(x+L\hat{\nu})  =  \Omega_\nu U_\mu(x) \Omega_\nu^{-1}\;,
\textrm{\hspace{2cm}}
\Psi(x+L\hat{\nu})  =  \Omega_\nu \Psi(x) \Omega_\nu^{-1}\;,
\end{equation}
where the quark field $\Psi_{sc}(x)$ becomes a matrix in smell-colour
space
\cite{Parisi:1984cy}
by the introduction of a new SU(N) quantum number ``smell''
in addition to the quark colour. We apply periodic boundary
conditions in the $(z,t)$ directions.

These boundary conditions lead to a change in the Fourier expansion of
the fields: in the twisted $(x,y)$ directions the momentum sums are
now over
\begin{equation}
p_\nu = m n_\nu,~~-\frac{NL}{2} < n_\nu \le \frac{NL}{2},~~\nu = (x,y)\;,
\end{equation}
where $m = \frac{2\pi}{N L}$.
The modes with ($n_x=n_y=0 \textrm{ mod } N$) are omitted from the sum
in the case of the gluons. The momentum sums for quark loops
need to be divided by $N$ to remove the redundant smell factor.

The twisted theory can be viewed as a two-dimensional Kaluza-Klein theory
in the $(z,t)$ plane. Denoting $\mathbf{n}=(n_x,n_y)$, the
stable particles in the $(z,t)$ continuum limit of this effective
theory are called the A mesons ($\mathbf{n}=(1,0)$ or
$\mathbf{n}=(0,1)$) with mass $m$ and the B mesons
($\mathbf{n}=(1,1)$) with mass $\sqrt{2}m$
\cite{Snippe:1997ru}.
%

%%%%%%%%%%%%%%%%%%%%%%%%%%%%%%%%%%%%%%%%%%%%%%%%%%%%%%%%%%%%%%%%%%%%%%%%%%%%

\section{Small-mass expansions}

Although we ultimately wish to extrapolate to the chiral limit,
we cannot set $m_qa=0$ straight away, since the
correct chiral limit is $m_qa \to 0,~ma \to 0,~m_q/m > C$, where
$C$ is a constant determined by the requirement that a Wick rotation
can be performed without encountering a pinch singularity.

Therefore, we first expand some observable quantity $Q$ in powers
of $ma$ at fixed $m_qa$:
\begin{equation}\label{eqn:fit_in_ma}
Q(ma,m_qa)=a^{(Q)}_0(m_qa) + a^{(Q)}_2(m_qa) (ma)^2 +
\mathcal{O}\left((ma)^4,(ma)^4\log(ma)\right)
\end{equation}
where the coefficients in the expansion are all functions of $m_qa$.
There is no term at $\mathcal{O}\left((ma)^2\log(ma)\right)$ since the
gluon action is improved at tree-level to $O(a^2)$
\cite{Snippe:1997ru}.
Then, we expand the coefficients $a^{(Q)}_0(m_qa)$ in power of $m_qa$.

For $a^{(Q)}_0(m_qa)$ we have
\begin{equation}\label{eqn:fit0_in_mqa}
a^{(Q)}_0(m_qa)~=~b^{(Q)}_{0,0}\log(m_qa) + a^{(Q)}_{0,0}\;.
\end{equation}
Since we expect a well-defined continuum limit, $a^{(Q)}_0(m_qa)$
cannot contain any negative powers of $m_qa$, but may contain
logarithms; $b^{(Q)}_{0,0}$ is the anomalous dimension associated
with $Q$, and can be determined by a continuum calculation.

For $a^{(Q)}_2(m_qa)$ we find
\begin{equation}\label{eqn:fit2_in_mqa}
a^{(Q)}_2(m_qa)~=~\frac{a^{(Q)}_{2,-2}}{(m_qa)^2} + a^{(Q)}_{2,0}
+ \left(a^{(Q)}_{2,2} + b^{(Q)}_{2,2}\log(m_qa)\right)(m_qa)^2 +
   \mathcal{O}\left((m_qa)^4\right)\;.\nonumber
\end{equation}
After multiplication by $(ma)^2$, the $(m_qa)^{-2}$ contribution gives
rise to a continuum contribution to $Q$, and $a^{(Q)}_{2,-2}$ is
calculable in continuum perturbation theory. There can be no term in
$(m_qa)^{-2}\log(m_qa)$ since this would be a volume-dependent further
contribution to the anomalous dimension of $Q$, and there can be no
term in $\log(m_qa)$ since the action is tree-level $O(a^2)$ improved.

In the chiral limit $m_q\to 0$, the term $w_i$ that appears on the
right-hand side of Eqn.
(\ref{eqn:impcond_generic})
is $a^{(Q)}_{2,0}$.

%%%%%%%%%%%%%%%%%%%%%%%%%%%%%%%%%%%%%%%%%%%%%%%%%%%%%%%%%%%%%%%%%%%%%%%%%%%

\section{Twisted spectral quantities}

\begin{figure}
\begin{center}
\includegraphics[height=6.4cm,keepaspectratio=,clip=]{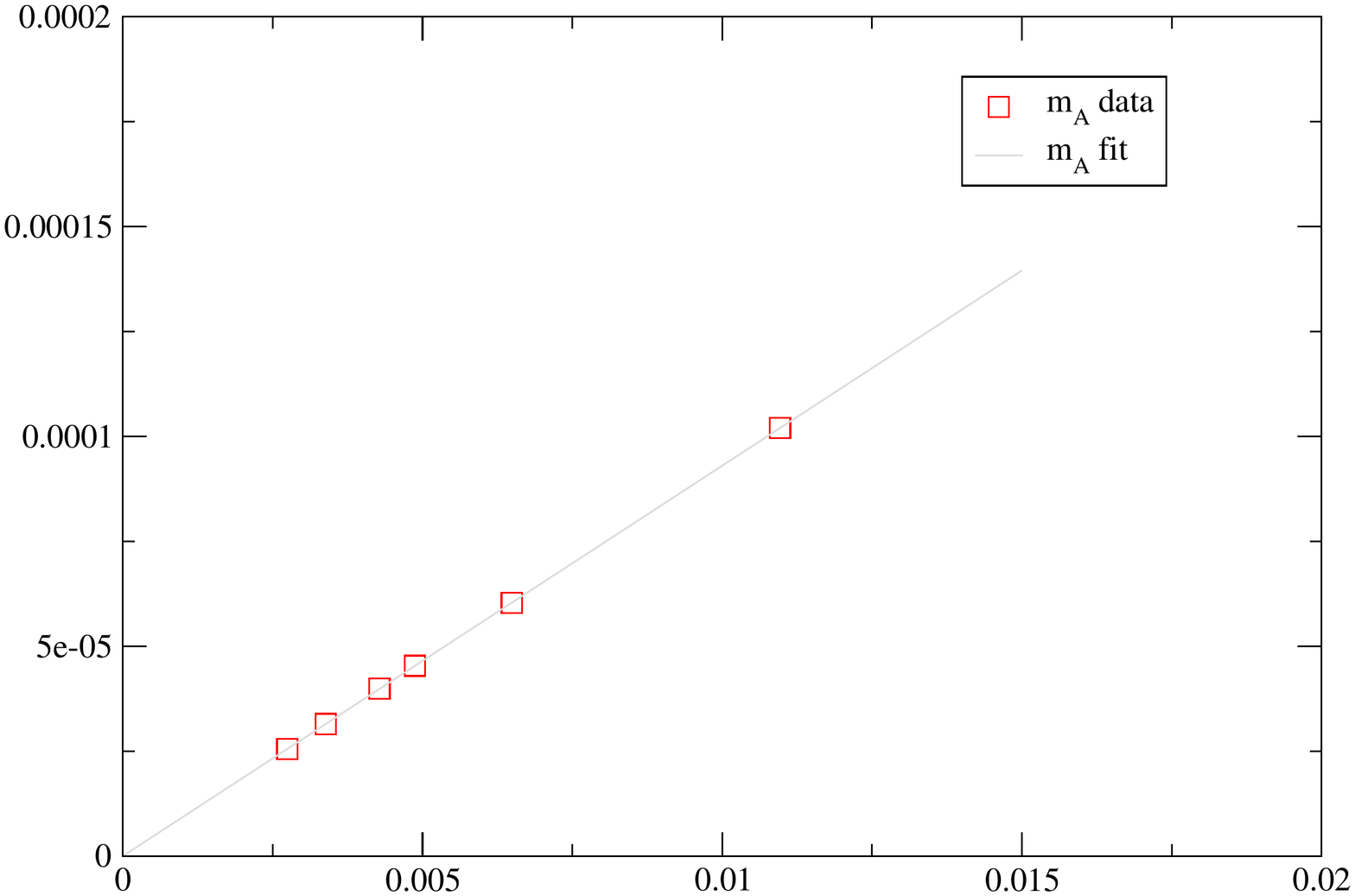}
\end{center}
\caption{A plot of the fermionic contributions to the one-loop $A$
  meson self-energy $m_A^{(1)}/m$ against $(ma)^2$. The vanishing of
  $m_A^{(1)}/m$ in the infinite-volume limit can be seen clearly.}
\end{figure}

The simplest spectral quantity that can be chosen within the framework
of the twisted boundary conditions outlined above is the
(renormalised) mass of the A meson. The one-loop correction the the A
meson mass is given by
\begin{equation}\label{eqn:mA}
m_A^{(1)} = - Z_0(\mathbf{k})
              \left.\frac{\pi_{11}^{(1)}(k)}{2 m_A^{(0)}}\right|
              _{k=(i m_A^{(0)},0,m,0)}
\end{equation}
where $Z_0(\mathbf{k})=1+\mathcal{O}\left((ma)^4\right)$ is the
residue of the pole of the tree-level gluon propagator at spatial
momentum $\mathbf{k}$, and $m_A^{(0)}$ is defined so that the momentum
$k$ is on-shell. 

From gauge invariance we find $a^{(m_A,1)}_{2,-2}~=~0$
and $a^{(m_A,1)}_0(m_qa)=0$.
The $\mathcal{O}\left(\alpha_s (ma)^2\right)$ contribution from
improvement of the action is given by
\cite{Snippe:1997ru}
\begin{equation}
\Delta_\textrm{imp} \frac{m_A^{(1)}}{m} = 
- ( c_1^{(1)} - c_2^{(1)} ) (ma)^2 + \mathcal{O}\left((ma)^4\right)\;.
\end{equation}

The next simplest independent spectral quantity is the
scattering amplitude for A mesons at B meson threshold, which can be
described by an effective $AAB$ meson coupling constant $\lambda$
\cite{Luscher:1985zq}:
\begin{equation}
\label{eqn:def_of_lambda}
\lambda =
g_0 \sqrt{ Z(\mathbf{k}) Z(\mathbf{p}) Z(\mathbf{q}) }
e_j \Gamma^{1,2,j}(k,p,q)
\end{equation}
with a twist factor of $\frac{i}{N}\mathrm{Tr}([\Gamma_k,\Gamma_p]\Gamma_q)$
factored out from from both sides,
and the momenta and polarisations of the incoming particles are
(where $r>0$ is defined such that $E(\mathbf{q})=0$)
\begin{equation}\label{eqn:momenta}
\begin{array}{llll}
k = (iE(\mathbf{k}),\mathbf{k}) \;\;\;\; &
p = (-iE(\mathbf{p}),\mathbf{p}) \;\;\;\;&
q = (0,\mathbf{q}) &  e = (0,1,-1,0) \\
\mathbf{k} = (0,m,ir)&\mathbf{p} = (m,0,ir)\;&\mathbf{q} = (-m,-m,-2ir)&
\end{array}
\end{equation}
We expand Eqn.
(\ref{eqn:def_of_lambda})
perturbatively to one-loop order and find (up to $\mathcal{O}((ma)^4)$
corrections)
\begin{equation}
\frac{\lambda^{(1)}}{m} =
\left( 1 - \frac{1}{24} m^2 \right)\frac{\Gamma^{(1)}}{m}
- \frac{4}{k_0} \frac{d}{dk_0}
           \left.\pi_{11}^{(1)}(k)\right|_{k_0=iE(\mathbf{k})}
 - \left( 1 - \frac{1}{12} m^2 \right) \frac{d^2}{dq_0^2}
           \left. \left( e^i e^j \pi_{ij}^{(1)}(q) \right) \right|_{q_0=0}
\end{equation}
The derivatives of Feynman diagrams are computed using automatic
differentiation
\cite{vonHippel:2005dh}.
Continuum calculations of the anomalous dimension and infrared
divergence give
\begin{equation}\label{eqn:anom_dir_lambda}
b^{(\lambda,1)}_{0,0}~=~-\frac{N_f}{3\pi^2}g^2\;,\;\;\;\;\;\
a^{(\lambda,1)}_{2,-2}~=~-\frac{N_f}{120\pi^2}g^2\;.
\end{equation} 
The improvement contribution to $\lambda$ is
\cite{Snippe:1997ru}
\begin{equation}
\Delta_\textrm{imp} \frac{\lambda^{1}}{m} = 4 (9 c_1^{(1)} - 7 c_2^{(1)}) (ma)^2
   + \mathcal{O}\left((ma)^4\right)\;.
\end{equation}

%%%%%%%%%%%%%%%%%%%%%%%%%%%%%%%%%%%%%%%%%%%%%%%%%%%%%%%%%%%%%%%%%%%%%%%%%%%

\section{Results}

\begin{figure}
\includegraphics[height=6cm,keepaspectratio=,clip=]{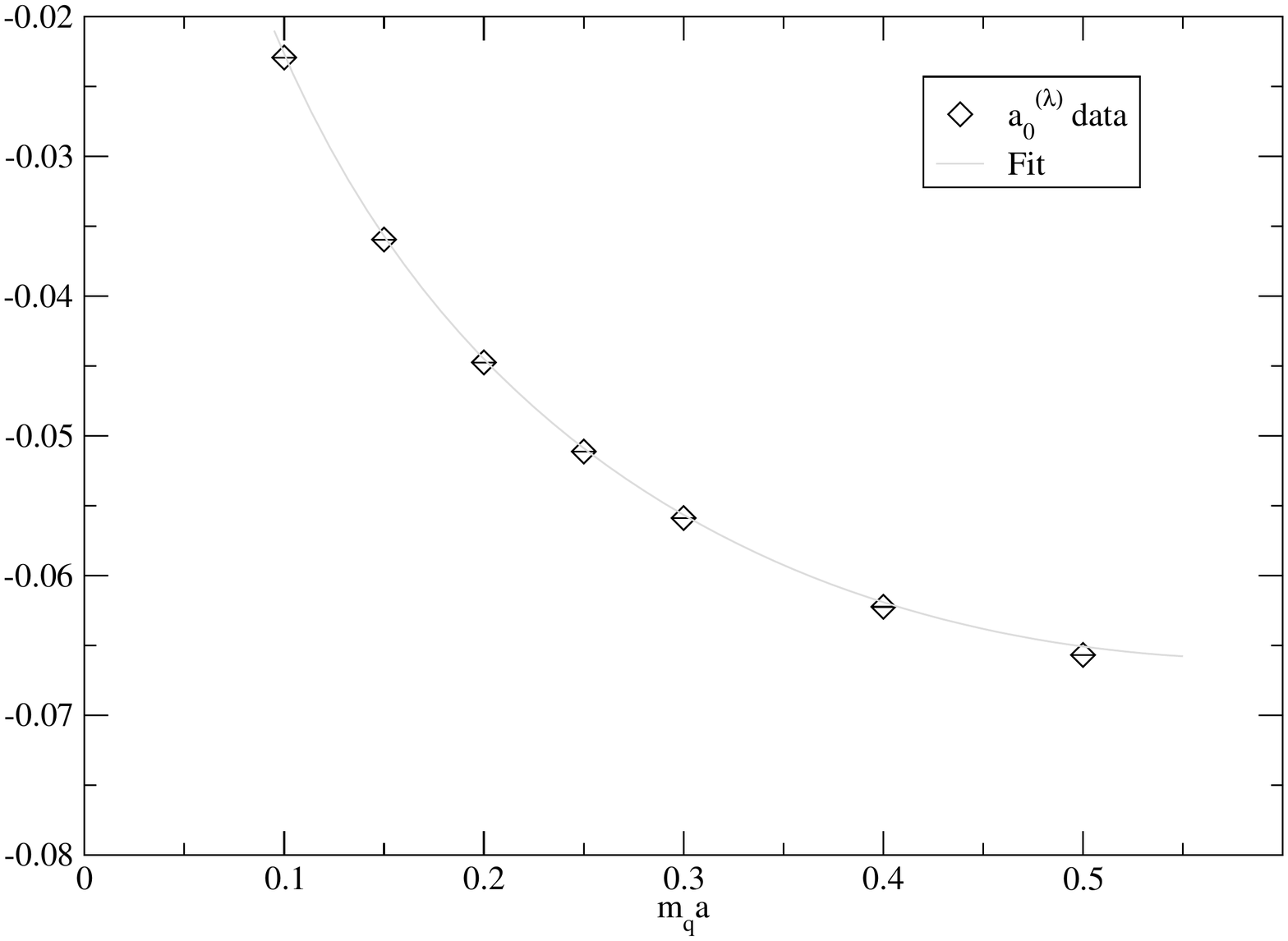}
\hfill
\includegraphics[height=6cm,keepaspectratio=,clip=]{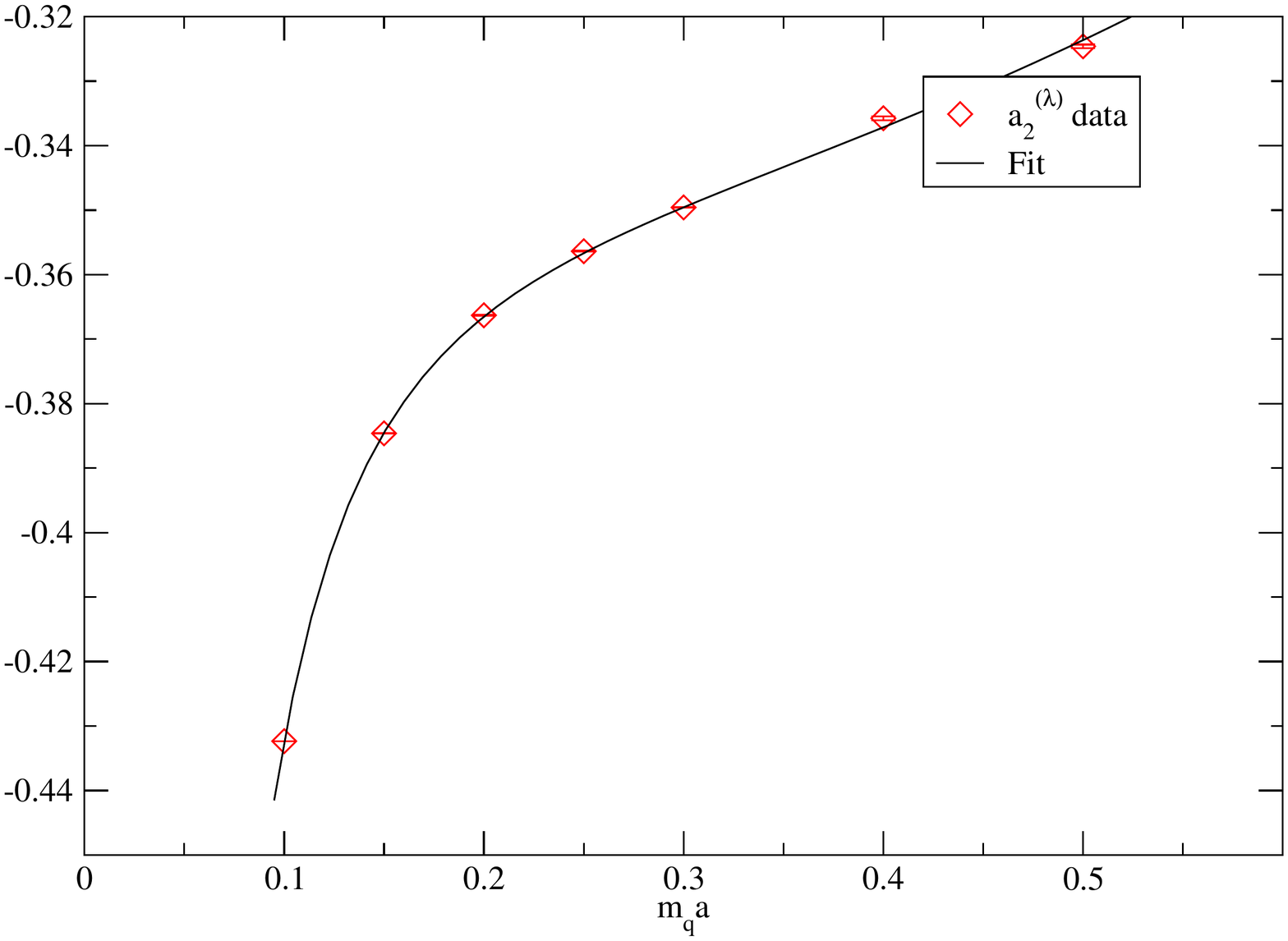}
\caption{
  Plots of $a_0^{(\lambda,1)}$ against $m_qa$ (left)
  and of $a_2^{(\lambda,1)}$ against $m_qa$ (right) with the fits
  shown for comparison.
  }
\end{figure}

To extract the improvement coefficients from our diagrammatic
calculations, we compute the diagrams for a number of different values
of both $L$ and $m_q$ with $N_f=1$, $N=3$. At each value of $m_q$, we
then perform a fit in $ma$ of the form given in Eqn.
(\ref{eqn:fit_in_ma})
to extract the coefficients $a_n^{(Q,1)}(m_qa),~n=0,2$. Our fits
confirm that $a_0^{(m_A,1)}(m_qa)=0$.

Performing a fit of the form
(\ref{eqn:fit0_in_mqa}) and (\ref{eqn:fit2_in_mqa}),
respectively, on these coefficients, we are able to extract the
analytically-known coefficients with high accuracy, along with the
required $(ma)^2$ contributions.

Solving equation
(\ref{eqn:impcond_generic})
for $c_i^{(1)}$ given the fitted values for $a_{2,0}^{Q_i}$,
our results can be summarised as
\begin{eqnarray}
c_1^{(1)} & = & -0.025218(4) + 0.0110(3) N_f \\
c_2^{(1)} & = & -0.004418(4) + 0.0016(3) N_f
\end{eqnarray}
where the quenched ($N_f=0$) results are taken from
\cite{Snippe:1997ru}.
The shift from the unquenched values is surprisingly large, even compared
to the coefficients for asqtad fermions \cite{Hao:2007iz}.
At first sight, this may
seem like a surprise, since HISQ is supposed to be the more highly-improved
action. However, HISQ is designed to suppress taste-changing interactions
(low momentum quark/high momentum gluon couplings), but these coefficients
come from high momentum quark/low momentum gluon couplings, for whose
suppression the HISQ action is not tuned.

\section*{Acknowledgments}

AH thanks the U.K. Royal Society for financial support. GMvH was supported
by the Deutsche Forschungsgemeinschaft in the SFB/TR 09.

This work has made use of the resources provided by the Edinburgh Compute
and Data Facility (ECDF;
\href{http://www.ecdf.ed.ac.uk}{http://www.ecdf.ed.ac.uk}). The ECDF is
partially supported by the eDIKT initiative
(\href{http://www.edikt.org}{http://www.edikt.org}).

%%%%%%%%%%%%%%%%%%%%%%%%%%%%%%%%%%%%%%%%%%%%%%%%%%%%%%%%%%%%%%%%%%%%%%%%%%%


\begin{thebibliography}{99}

\bibitem{Davies:2003ik}
C.~T.~H. Davies {\em et~al.},
\newblock Phys. Rev. Lett. {\bf 92}, 022001 (2004)
 [\href{http://arxiv.org/abs/hep-lat/0304004}{{\tt hep-lat/0304004}}];
%%CITATION = HEP-LAT 0304004;%%
C. Aubin {\em et~al.},
\newblock Phys. Rev. {\bf D70}, 114501 (2004)
 [\href{http://arxiv.org/abs/hep-lat/0407028}{{\tt hep-lat/0407028}}];
%%CITATION = HEP-LAT 0407028;%%
E.~Follana, {\em et~al.},
\newblock Phys.\ Rev.\ Lett.\  {\bf 100}, 062002 (2008)
  [\href{http://arxiv.org/abs/0706.1726}{{\tt arXiv:0706.1726}}].
%%CITATION = PRLTA,100,062002;%%

\bibitem{Orginos:1999cr}
K.~Orginos, D.~Toussaint and R.~L. Sugar,
\newblock Phys. Rev. {\bf D60}, 054503 (1999)
 [\href{http://arxiv.org/abs/hep-lat/9903032}{{\tt hep-lat/9903032}}].
%%CITATION = HEP-LAT 9903032;%%

\bibitem{Sharpe:2006re}
S.~R.~Sharpe,
\newblock PoS {\bf LAT2006}, 022 (2006)
 [\href{http://arxiv.org/abs/hep-lat/0610094}{{\tt hep-lat/0610094}}].
%%CITATION = POSCI,LAT2006,022;%%

\bibitem{Follana:2006rc}
E.~Follana {\it et al.},
\newblock Phys. Rev. {\bf D75}, 054502 (2007)
 [\href{http://arxiv.org/abs/hep-lat/0610092}{{\tt hep-lat/0610092}}].
%%CITATION = PHRVA,D75,054502;%%

\bibitem{Wong:2007uz}
K.~Y.~Wong and R.~M.~Woloshyn,
\newblock PoS {\bf LAT2007}, 047 (2007)
  [\href{http://arxiv.org/abs/0710.0737}{{\tt arXiv:0710.0737}}].
%%CITATION = POSCI,LAT2007,047;%%

\bibitem{Hao:2007iz}
Zh.~Hao, G.~M.~von Hippel, R.~R.~Horgan, Q.~J.~Mason and H.~D.~Trottier,
\newblock Phys. Rev. {\bf D76}, 034507 (2007)
 [\href{http://arxiv.org/abs/0705.4660}{{\tt arXiv:0705.4660}}].
%%CITATION = PHRVA,D76,034507;%%

\bibitem{Hart:2004bd}
A.~Hart, G.~M. von Hippel, R.~R. Horgan and L.~C. Storoni,
\newblock J. Comput. Phys. {\bf 209}, 340 (2005)
 [\href{http://arxiv.org/abs/hep-lat/0411026}{{\tt hep-lat/0411026}}].
%%CITATION = HEP-LAT 0411026;%%

\bibitem{Luscher:1984xn}
M.~L\"uscher and P.~Weisz,
\newblock Commun. Math. Phys. {\bf 97}, 59 (1985).
%%CITATION = CMPHA,97,59;%%

\bibitem{Luscher:1985wf}
M.~L\"uscher and P.~Weisz,
\newblock Nucl. Phys. {\bf B266}, 309 (1986).
%%CITATION = NUPHA,B266,309;%%

\bibitem{Snippe:1997ru}
J.~Snippe,
\newblock Nucl. Phys. {\bf B498}, 347 (1997)
 [\href{http://arxiv.org/abs/hep-lat/9701002}{{\tt hep-lat/9701002}}].
%%CITATION = HEP-LAT 9701002;%%

\bibitem{Parisi:1984cy}
G.~Parisi,
\newblock Invited talk given at Summer Inst. Progress in Gauge Field Theory,
  Cargese, France, 1983.

\bibitem{Luscher:1985zq}
M.~L\"uscher and P.~Weisz,
\newblock Phys. Lett. {\bf B158}, 250 (1985).
%%CITATION = PHLTA,B158,250;%%

\bibitem{vonHippel:2005dh}
G.~M.~von Hippel,
\newblock Comput.\ Phys.\ Commun.\  {\bf 174}, 569 (2006)
  [\href{http://arxiv.org/abs/physics/0506222}{{\tt physics/0506222}}].
%%CITATION = CPHCB,174,569;%%


\end{thebibliography}
\end{document}